# Beyond the Newtonian Paradigm:

# A Statistical Mechanics of Emergence

Stuart A. Kauffman[1,2] and Andrea Roli[3,4]

[1] Institute for Systems Biology, Seattle, USA
[2] Emeritus Professor of Biophysics and Biochemistry, University of Pennsylvania
[3]Department of Computer Science and Engineering, Campus of Cesena,
*Alma Mater Studiorum* Università di Bologna
[4]European Centre for Living Technology, Venezia, Italy

**Abstract**

Since Newton, all classical and quantum physics depends upon the "Newtonian Paradigm". Here the relevant variables of the system are identified. For example, we identify the position and momentum of classical particles. Laws of motion in differential form connecting the variables are formulated. An example is Newton's three Laws of Motion and Law of Gravitation. The boundary conditions creating the phase space of all possible values of the variables are defined. Then, given any initial condition, the differential equations of motion are integrated to yield an entailed trajectory in the pre-stated and fixed phase space. It is fundamental to the Newtonian Paradigm that the set of possibilities that constitute the phase space is always definable and fixed ahead of time.

All of this fails for the diachronic evolution of ever new adaptations in our, or any biosphere. The central reason is that living cells achieve Constraint Closure and construct themselves. With this, living cells, evolving via heritable variation and Natural selection, adaptively construct new in the universe possibilities. The new possibilities are opportunities for new adaptations thereafter seized by heritable variation and Natural Selection. Surprisingly, we can neither define nor deduce the evolving phase spaces ahead of time. The reason we cannot deduce the ever evolving phase spaces of life is that we can use no mathematics based on Set Theory to do so. We can neither write nor solve differential equations for the diachronic evolution of ever new adaptations in a biosphere.

These ever-new adaptations with ever-new relevant variables constitute the ever-changing phase space of evolving biospheres. Because of this, evolving biospheres are entirely outside the Newtonian Paradigm.

One consequence is that for any universe such as ours with one or more evolving biospheres, there can be no Final Theory that entails all that comes to exist. The implications are large. We face a third major transition in science beyond the Pythagorean dream that "All is Number", a view echoed by Newtonian physics.

In face of this, we must give up deducing the diachronic evolution of the biosphere. All of physics, classical and quantum, however, apply to the analysis of existing life, a synchronic analysis.

But there is much more. We begin to better understand the emergent creativity of an evolving biosphere. Thus, we are on the edge of inventing a physics-like new statistical mechanics of emergence.

Wholes, Abstract Kantian Wholes, Functional Closure, Constraint Closure, Emergent Creativity, Statistical Mechanics of Emergence.

# 1 Introduction

Three centuries after Newton we are, we believe, at a third major transition in science. We hope to make clear the evidence and need for this transition, and the wide, unexpected landscape for new science that can be glimpsed.

We may attribute the first major transition to Newton, the invention of the differential and integral calculus, and the invention of classical physics. It is no understatement that Newton taught us how to think. Call this "The Newtonian Paradigm" (Smolin, 2013): First, find the relevant variables. In physics these are often position and momentum. Write laws of motion for these relevant variables in ordinary or partial differential deterministic equation form, or stochastic variants. Define ahead of time the boundary conditions, hence the phase space of all possible values of the relevant variables such as positions and momenta of particles of the system. For any initial condition of the relevant variables, integrate the laws of motion to obtain the entailed trajectory of the system in its phase space. It is fundamental to the Newtonian Paradigm that we can and must always define the phase space ahead of time.

Classical physics including General Relativity gave us the "clockwork" universe that will unfold deterministically with a Deistic God no longer able to work miracles. The same clockwork universe renders "chance" merely epistemic. And it renders "mind" hapless at best.

The second major transition is nothing less than the reluctant discovery of the quantum of action in 1900 (Planck, 1901), thence the miracles of quantum mechanics and quantum field theory (Heisenberg, 1958; Feynman, 1998). Quantum theory, however, remains safely within the Newtonian Paradigm with a prestated phase space, including Fock space, hence initial and boundary conditions, and the deterministic evolution of a probability distribution via the Schr'odinger wave equation. Determinism is broken, on most interpretations of quantum mechanics, on the Born rule and von Neumann's projection postulate (Birkhoff and Von Neumann, 1936). Among the most astonishing implications is spatial non-locality (Einstein et al., 1935; Aspect et al., 1982).

The enormous power of the Newtonian Paradigm can be found outside of physics. Ecology often considers a community of species linked by non-linear dynamical equations of motion concerning the rate of reproduction of members of each species and the food web among the species. Integration of the equations in the predefined phase space of the relevant variables may exhibit limit cycles, multiple attractors, and other aspects of non-linear dynamical systems (Svirezhev, 2008).

The foundational theory of microeconomics, Competitive General Equilibrium, CGE (Arrow and Debreu, 1954), is firmly within the Newtonian Paradigm. CGE addresses the problem of the existence of an "equilibrium" vector of prices among a set of goods such that the supply and demand for all goods is balanced and "markets clear". Market clearing is the concept of "equilibrium". Consider the well-known supply and demand curve for a single good, such as bread. Supply and demand are inversely related to the price of the good. At the equilibrium price, supply equals demand and all supplied goods are sold. The market clears. The problem arises for two goods that are used together, bread and butter. If the price of butter goes up, the demand for bread will go down. Does an equilibrium pair of prices for bread and butter exist? For an arbitrarily large number of goods, will a vector of prices exist such that all markets for these goods clear? In short does an equilibrium exist?

Arrow and Debreu solved the problem in 1954. We are to consider “All possible dated-contingent goods”. An example of a dated contingent good is “a kilogram of apples on your door-step only if it rains in Shanghai on March 15th of this year”. Next, we suppose that all the economic actors are infinitely rational and they also have probabilities, or expectations, with respect to all possible dated contingent goods. And all agents also have their own utility function, or preferences. At the beginning of time, an auctioneer auctions off contacts for all possible dated-contingent goods. Contracts are let. In this setting Arrow and Debreu prove a fixed-point theorem in this continuum of dated-contingent goods showing that at least one vector of prices exists such that no matter how the future unfolds, all markets clear. It is a remarkable result. Competitive General Equilibrium remains within the Newtonian Paradigm. The prestated and also fixed phase space is the continuum of “All possible dated-contingent goods”.

CGE remains the foundation of microeconomics. There are familiar doubts about additional “sunspot” equilibria (Cass and Shell, 1983), and the implications of incomplete markets, incomplete knowledge, and other issues (Geanakoplos and Polemarchakis, 1986). We now wish to place ecology and CGE in a wider context. Ecology deals with a predefined set of species in a community. These provide the relevant variables, hence the predefined phase space. Over evolutionary time species come and go. The set of species and their patterns of interactions themselves evolve. In the diachronic evolution of the biosphere, new adaptations emerge, existing adaptations vanish by extinction. Ecology can hope to be valid over time scales such that the species do not evolve relevant new features or lose relevant old ones. The issue we wish to raise, and the central question of this article, asks whether we can predict or deduce the new relevant adaptive variables that arise and old ones that vanish. Can we have well founded expectations? We hope to demonstrate that the answer is “No”.

The same issue arises with respect to CGE. We are asked to consider all possible dated-contingent goods and have well formulated expectations with respect to them. But over the past 50,000 years of diachronic evolution of the econosphere, the number of goods has exploded from a few thousand to billions today. New goods arise, old goods vanish. The issue we again wish to raise, central to this article, asks whether we can predict or deduce the new relevant economic variables that arise and old ones that vanish? Can we have well founded expectations? We hope again to demonstrate that the answer is “No”.

If we cannot deduce the ever-changing phase space it will be because we will be unable to write or solve equations of motion allowing deduction of those changing phase spaces. We will be outside of the Newtonian Paradigm.

Life on earth has existed for almost 4 billion years, almost 30% of the lifetime of the universe. A failure of the Newtonian paradigm with respect to evolving life, let alone the evolving econosphere, will mean that major aspects of the cosmological evolution of the universe are outside of the Newtonian Paradigm.

## 2 The Non-Deducible Diachronic Evolution of the Biosphere

Life started on earth about 3.7 billion years ago. The biosphere is the most complex system we know in the universe. The central new issue is that it really is not possible to deduce the diachronic evolution of our or any biosphere. The evolving biosphere is a propagating construction not an entailed deduction (Montévil and Mossio, 2015; Kauffman, 2020; Longo et al., 2012).

The reasons seem, at first, strange (Kauffman and Roli, 2021):

**1)** The universe is not ergodic above the level of about 500 Daltons (Kauffman et al., 2020). The universe really will not make all possible complex molecules such as proteins 200 amino acids long in vastly longer than the lifetime of the universe (Kauffman, 2019; Cortes et al., 2021). Because the universe is not ergodic on time scales very much longer than the lifetime of the universe, it is true that most complex things will never "get to exist".

**2)** Human hearts, very complex things weighing 300 grams and able to function to pump blood, exist in the universe. How can that be possible? The fundamental answer for why hearts exist in the universe is that life, based on physics, arose, evolved, and adapted in that evolution over time. Living things have a special organization of non-equilibrium processes. Living things are Kantian Wholes where the Parts exist in the universe for and by means of the Whole. Humans are Kantian Wholes. We exist for and by means of our parts, such as hearts pumping blood, and kidneys purifying the blood in the loops of Henle making and excreting urine. Because we, as Kantian Wholes, propagate our offspring, our sustaining parts, hearts and kidneys are also propagated and evolve to function better. The "function" of the heart is to pump blood, not jiggle water in the pericardial sac. The function of a Part is that subset of its causal properties that sustains the Whole (Kauffman, 2019).

**3)** We cannot hope to account for the existence in the universe of a heart that can pump blood, or the loop of Henle in the kidney that can purify urine, without appeal to the function of these organs and their adaptive diachronic evolution by Darwin's heritable variation and Natural Selection. Selection is downward causation. Selection acts on the whole organism, not its evolving parts. What gets to exist in the evolving biosphere is that which was selected. The explanatory arrows point upward. The selection of the whole alters the parts.

**4)** In more detail, a Kantian Whole has the property that the Parts exist for and by means of the Whole. A simple physical example is an existing 9 peptide collectively autocatalytic set (Kauffman, 2019; Ashkenasy et al., 2004). Here peptide 1 catalyzes a reaction forming a second copy of peptide 2 by ligating half fragments of peptide 2 into a second copy of peptide 2. The half fragments are "food" fed from an exogenous source. Similarly, peptide 2 catalyzes the formation of a second copy of peptide 3. And so on around a cycle in which peptide 9 catalyzes a second copy of peptide 1. The entire set of nine peptides is collectively autocatalytic. The set is a Kantian Whole.
This collectively autocatalytic physical set has these properties:

1. It is collectively autocatalytic (Ashkenasy et al., 2004). No molecule catalyzes its own formation. Thus, this is a Kantian Whole, the Parts do exist for and by means of the Whole.
2. The function of a Part is that subset of its causal properties that sustains the Whole. The function of peptide 1 is to catalyze the formation of a second copy of peptide 2. If, in doing so, the peptide jiggles the water in the Petri plate, that causal consequence is not the function of peptide 1.
3. The system achieves Catalytic Closure: All reactions needing catalysis have catalysts within the same system.

4. The system achieves the newly recognized and powerful property of Constraint Closure (Montévil and Mossio, 2015): Thermodynamic work is the constrained release of energy into a few degrees of freedom (Atkins, 1984). These constraints constitute boundary conditions. The peptides in the nine peptide collectively autocatalytic set are each a physical boundary condition that constrains the release of chemical energy. Each peptide binds the two substrates of the next peptide, thus lower activation barrier, thus chemical energy is released into a few degrees of freedom, and thermodynamic work is done to ligate the two fragments and construct the next peptide. Critically, the set of peptides construct themselves, thus construct the very constraints on the release of energy that constitutes the work by which they construct themselves! This is Constraint Closure (Montévil and Mossio, 2015; Kauffman, 2019, 2020).

Cells literally construct themselves. The evolving biosphere constructs itself. Automobiles do not construct themselves. We construct our artefacts. Living cells constitute a new class of matter and organization of process that is a new union of thermodynamic work, catalytic closure and constraint closure (Montévil and Mossio, 2015). In a real sense this is the long sought "vital force", here rendered entirely non-mystical.

It is critical to emphasize that because living cells are open thermodynamic systems that construct themselves, they can and do construct ever new boundary conditions that thereby create new in the universe phase space possibilities that were not prestated (Kauffman, 2020). Not only do the boundary conditions change, but ever new "relevant variables" emerge and constitute the new phase space. For example, with respect to the heart, systolic blood pressure, diastolic blood pressure, cardiac blood ejection volume, and blood oxygenation are among the now functionally relevant variables. Consider mimicry among butterflies. Good tasting butterflies have evolved wing color patterns to mimic bad tasting butterflies as camouflage to avoid predation by birds. The newly relevant variables for the butterflies involves the recognition capacities of the birds and the specific features of the bad tasting butterflies. How are we to account for this without adaptive evolution of ever novel functionalities?

**5)** The spontaneous emergence of life, of molecular Kantian Wholes, in the evolving universe may well be an expected phase transition in complex chemical reaction networks. This body of theory and experiments is part of a theory of the origin of life on earth and elsewhere that has developed over the past 50 years (Kauffman, 1971, 1986; Farmer et al., 1986; von Kiedrowski, 1986; Lincoln and Joyce, 2009; Vaidya et al., 2012; Hordijk and Steel, 2004; Lancet et al., 2018; Xavier et al., 2020). The central idea is a phase transition to collectively autocatalytic sets in sufficiently complex chemical reaction networks. Experimental collectively autocatalytic sets comprised of DNA, of RNA, and of peptides have been constructed. Most astonishingly, Xavier et al. Xavier et al. (2020) analyzed Archea and bacteria from before oxygen was in the atmosphere and found in each a small molecule collectively autocatalytic set containing no polymers at all. No DNA, no RNA, no proteins.

Even more wonderfully, the small molecule autocatalytic sets in Archea and bacteria overlap in an intersection subset of 172 reactions and small molecules that is itself collectively autocatalytic. This strongly suggests the smaller intersection subset was present in LUCA, the ancestor of Archea and bacteria before the two kingdoms of life diverged. These are molecular fossils from more than 2 billion years ago.

These observations very strongly suggest that life arose as small molecule collectively autocatalytic sets, very plausibly as the phase transition proposed. By 4 or 5 billion years ago the universe had cooked up a high diversity of small molecules, as seen in the Murchison meteorite formed with the solar system. This meteorite has tens of thousands of organic molecules (Kvenvolden et al., 1970). If such a diversity easily yields the spontaneous formation of small molecule collectively autocatalytic sets, life is abundant among the solar systems in the universe.

On the earth, an early formation of small molecule collectively autocatalytic sets that synthesize, as observed (Xavier et al., 2020), amino acids and nucleotides may well have supported the subsequent formation of peptide-RNA autocatalytic sets that also evolved to catalyze the reactions already present in the small molecule autocatalytic metabolism that sustained the emerging peptide-RNA system (Lehman and Kauffman, 2021). This emergence of early life would be followed by template replication and coding (Lehman and Kauffman, 2021). All this is now testable experimentally. Once such life arose it was a Kantian Whole achieving collective catalysis and Constraint Closure. Even without genes, such systems can evolve to some extent so acquire new adaptations (Vasas et al., 2012). With the emergence of coding, that mystery, evolving life is fully formed on earth.

The alternate and standard view of the origin of life posits the emergence of at least one template replicating RNA sequence able to copy itself (Joyce, 2002). This has not yet been experimentally successful but may well become demonstrated. This theory faces the issue that ribonucleotides and polymers of RNA are hard to synthesize on the early earth. Moreover, any such "nude" replicating RNA would need to evolve ribozymes to catalyze some connected metabolism that can sustain the RNA polymer system. This is easy to imagine. However there seems to be no reason at all why such a connected small molecule metabolism should itself be collectively autocatalytic. Why would such a de novo metabolism be able to reproduce itself? That new catalyzed metabolism was selected merely to sustain the RNA world system that uses it.

These facts now almost persuasively indicate that life arose as small molecule collectively autocatalytic sets, perhaps widely in the universe. If life is widely distributed among the solar systems in the universe and that evolution is beyond the Newtonian paradigm, vast new domains of science must be created with respect to major aspects of the evolving universe.

**6)** Most adaptations in the evolution of the biosphere are "affordances", typically seized by heritable variation and Natural Selection. An example of an affordance (Gibson, 1966) is a horizontal surface which affords you a place to sit. Affordances are, in general "The possible use by me of X to accomplish Y". "Accomplish" can occur without "mind", but by "blind" heritable variation and natural selection, as in the evolution of the heart and loop of Henle (Kauffman and Roli, 2021).

An affordance is not an independent feature of the world (Walsh, 2015). An affordance is in relation to the evolving organism for whom it is an affordance to be seized or not by heritable variation and natural selection. Biological degrees of freedom are affordances, or relational opportunities available to evolving organisms.

**7)** Often in evolution adaptations emerge by co-opting the same organ for a new function. These are called Darwinian preadaptations or exaptations (Gould and Vrba, 1982).

Typical examples of such an affordance, or new Darwinian preadaptation, seized by heritable variation and natural selection include flight feathers, which evolved earlier for functions such as thermal insulation or as bristles but were co-opted for the new function of flight (Prum and Brush, 2002; Persons IV and Currie, 2015), and lens crystallins originated as enzymes (Barve and Wagner, 2013). A wonderful example is the evolution of the swim bladder that emerged in a lineage of fish (Kauffman, 2016). In this latter instance, the ratio of air and water in the swim bladder functions to assess neutral buoyance in the water column. Paleontologists believe the swim bladder arose from the lungs of lung fish. Water got into some lung, now a sac filled with a mixture of air and water, so poised to evolve into a swim bladder. This is precisely finding a new use for the same initial "thing", the lung. A new function, neutral buoyancy in the water column, has emerged in the evolving biosphere. There is yet more: Once a swim bladder emerged it becomes newly possible that a worm or bacterium might evolve to live in swim bladders. Natural Selection presumably "worked" to craft a functioning swim bladder. But did Natural Selection "craft" the swim bladder such that it could become a new affordance, a new niche which might be seized by the worm or bacteria? No. Without Selection "achieving it", the evolving biosphere is creating the ever-new affordances, the ever-new niche possibilities, into which the biosphere evolves. The biosphere is constructing the very Adjacent Possible into which it enters (Kauffman, 2019).

# 3 The Insuperable Limits of Set Theory

We have established that in the evolution of the biosphere, ever new phase spaces with new boundary conditions and new relevant variables arise that were not prestated. Then it is essential to ask if those now relevant variables might have been prestated. The surprising answer, we hope to show, is "NO".

We cannot prestate the ever-new relevant variables because we can neither compute, predict, nor deduce ahead of time the coming into existence of new affordances and newly relevant variables seized by heritable variation and natural selection.

We cannot compute or deduce the new adaptive phase spaces because we cannot use Set Theory or any mathematics based on it to reliably and soundly model the evolutionary emergence of adaptations as "seized affordances". The considerations are a bit unexpected and focus on the implications of biosphere evolution features for the foundations of Set Theory (Kauffman and Roli, 2021).

Although our argument concerns the case of affordances seized by evolution—and not cognitive ones—we believe an example from the tool usage context may be greatly explicative. How many "uses" does a screwdriver have, alone or with other things, in London on March 22, 2021? *i.* Screw in a screw. *ii.* Open a can of paint. *iii.* Wedge a door closed. *iv.* Scrape putty off a window. *v.* As an *objet d'art*. *vi.* Tie to a stick and spear a fish. *vii.* Rent the spear to local fishermen and take 5% of the catch. *viii.* Lean the screwdriver against a wall, place plywood propped up by the screwdriver and use this to shelter a wet oil painting, etc.

Is the number of uses of a screwdriver alone or with other things a specific number, say 11? No. Is the number of uses infinite? How would we know? The number of uses of a screwdriver now and over the next 1000 years is "*indefinite*" or perhaps "*unknown*". No one in 1690 could have used a screwdriver to short an electric connection. It is essential to remark that we *cannot list all the possible uses of a screwdriver* (Kauffman, 2019) as not only can we not predict the possible future niches for the screwdriver, but the uses of a screwdriver also depends upon user's goals and repertoire of actions (Walsh, 2015). The same considerations apply in general to any object, e.g. to the uses of an engine block. It can be used to build an engine, as a chassis for a tractor, as a paper weight, to crack open coconuts against its sharp corners, etc.

One may argue that we cannot list all the possible uses of an object by using our intuition, but we could do it by applying enumeration or deduction. This is not possible either. There are four mathematical ordering scales, Nominal, Partial Order, Interval, Ratio. The uses of an object are merely a nominal scale, therefore *there is no ordering relation between these uses*. Furthermore, in general a specific use of an object does not provide the basis for entailing another use. Hence, there is no *deductive relation* between the different uses of an object, e.g. it is not possible to deduce the use of an engine block to crack open coconuts from its use as a paper weight.

We believe it is apparent that these arguments hold also for the emergence of adaptations as seized affordances along the diachronic evolution of the biosphere: ever-new affordances appear, which are seized by evolution and shape ever-new niches and biological functions in an unpredictable way. Two main observations support this statement: first, the articulation of parts explanation (Kauffman, 1970) and the impredicative loop among affordances and agent's goals and actions (Walsh, 2015). In one sentence: biological evolution concerns constructing, not listing.

The implication of what we stated above is that *we cannot use Set Theory with respect to the diachronic emergence of new affordances seized by heritable variation and natural selection* (Kauffman and Roli, 2021). *This concerns all the adaptations in the diachronic evolution of the biosphere*.

A first axiom of Set Theory is the *axiom of extensionality*: "Two sets are identical if and only they contain the same members" (Jech, 2006). But we cannot prove that the *un-listable* uses of a screwdriver are identical to the *un-listable* uses of an engine block, as we cannot prove, once and for all, the uses of object *X*. Therefore, no axiom of extensionality. Hence, no sound Set Theory can be formulated.

Worse, the implications also reach mathematical fields based on Set Theory. The *axiom of choice* (Moore, 2012), which comes into play whenever a choice function cannot be defined, cannot be applied. The axiom of choice is equivalent to "well ordering" (Potter, 2004), but an ordering among the unordered uses of X cannot exist. Therefore, we would reach a contradiction if we tried to postulate this in a formal description of the evolution of affordances.

A consequence of this argument is the impossibility of using numbers with respect to the emergence of novel functions in the evolving biosphere. One way to define numbers uses Set Theory (Russell and Whitehead, 1973). The number "0" is defined as the set of all sets each of which has 0 elements. In our case this corresponds to "the set of all objects that have exactly 0 uses." Well, no, this cannot be grounded on objects in an evolving biosphere. The alternative approach to numbers is via Peano's Axioms (Peano, 1889). These require a null set and a successor relation. But we have no null set. More, the different uses of X are unordered. We have no successor relation.

Therefore, with respect to all diachronically emerging adaptations via seizing affordances, no numbers. No integers, no rational numbers, no equations such as 2+3=5. No equations so no irrational numbers. No real line. No equations with variables. No imaginary numbers, no quaternions, no octonions. No Cartesian spaces. No vector spaces. No Hilbert spaces. No union and intersection of uses of X and uses of Y. No first order logic. No combinatorics. No topology. No manifolds. No differential equations on manifolds.

Further, without an Axiom of Choice, we cannot integrate and take limits on the differential equations we cannot write.[1]

# 4 The Third Transition: We are Beyond the Newtonian Paradigm

These facts mean that we are, surprised or not, at the third major transition in science. If we can neither write nor solve differential equations for the diachronic evolution of adaptations in the biosphere we cannot, in principle, prestate, compute, or deduce the relevant ever-new phase spaces of evolving biospheres. The evolving biosphere advances into the adjacent possible it creates, but we cannot deduce what is "in" that adjacent possible. Therefore, we do not know the sample space of the process, hence can neither define a probability measure, nor define "random". We truly have no well-founded expectations. With respect to the diachronic evolution of new adaptations, we are beyond the Newtonian Paradigm.

The implications are very large. If we can write and solve no equations for the diachronic evolution of our or any biosphere and our evolving universe has at least one evolving biosphere, there can be no final theory that entails what comes to exist in the evolving universe. The famous equation destined for the T-shirt (Kaku, 2021), it now seems, does not exist.

This result is somewhat stunning at first, then perhaps not totally surprising. Gödel's First Incompleteness Theorem (Franzén, 2005) assures us that any consistent axiomatic system as rich as arithmetic has the property that, given the axioms and the inference rules, a statement exists such that it can neither be proved nor disproved inside the system. The non provable statement is itself generated algorithmically (Longo, 2019). If this algorithmically generated statement itself is added to the initial axioms, the new set of axioms again algorithmically generates statements whose truth cannot be neither proved nor disproved. *In short, Gödel's theorem, iterated, yields an open succession of ever new axiom systems. Gödel's theorem relies on self-reference*.

The evolving biosphere instantiates Gödel's Theorem, and even more. New adaptations, new uses of physical things such as molecules, as is true for the new uses of an engine block, cannot be deduced from the old uses. And importantly, affordances are referential to the organism for whom the affordance is relevant. Affordances are referential degrees of freedom, not independent features of the world. Thus, the referential new uses cannot be deduced as a theorem from knowledge of the properties and functions of the existing molecules and other physical properties of organisms prior to the new adaptation (Kauffman and Roli, 2021). Therefore, they are more than the analogue of algorithmically generated undecidable statements: They can be read as "If I get to exist in a new way for some time in the biosphere, my new existence cannot be deduced from the biosphere up to the present moment".

Like the ongoing generation of ever richer axiom systems in Gödel's Theorem by successive additions of new axioms, the generation of ever new non-deducible adaptations by new uses among the indefinite possible uses of each physical thing in an evolving cell or multicellular organism successively says, "I get to exist in a new way in the biosphere that is not deducible."

Reluctant or not, we observe that the evolution of our or any biosphere is outside of the Newtonian Paradigm. What are some implications?

[1] Both the (ε−δ) formal definition of limits (Grabiner, 1983) and the one based on *infinitesimals* (Robinson, 2016) rely on Set Theory.

1. There really can be no "final theory" that entails all that comes to exist in the evolving universe. The dream of such a final theory is magnificent and has been a driving motivation for superb science for centuries. Perhaps our arguments are wrong. If so, let them be vanquished.
2. The evolution of our or any biosphere in the universe is not only entailed by no law, but seems not even mathematizable by known techniques. Perhaps we can invent new mathematics.
3. If no law entails the evolution of biopsheres and that evolution cannot even be mathematized, biological evolution is radically "free" to be and is vastly creative. Section 7 below hopes to find some of the unexpected reason for such ongoing creativity.
4. Most essentially, we really are at a third transition in science. The scale and meanings of this are quite unclear at present. Our universe is creative in ways we have not known. Somehow our understanding of the world will change.

# 5 Toward A Statistical Mechanics of Emergence

There is a pathway forward. A beginning point is to realize that, in fact, any physical (or other) object has indefinitely many uses that are not deducible from one another. The engine block really can be used as a paper weight and to crack open coconuts. Therefore, we must give up specific "properties" of objects and abstract an object as just that, an "object" or "thing". Given this step, we can think of "things" transforming to yield old or new "things". And we can think of "things" acting on and regulating the transformations among "things" to yield old or new "things".

From such a spare abstraction, a great deal can already be done. Loreto et al. (Loreto et al., 2016) formulated a modified urn model. Here one starts with two "things" for example red and black balls in the urn. The process samples at random from the urn. If a new color, never seen before, is encountered, a ball with a new color is added to the urn. In this model, a single "thing" can only give rise to a single new "thing". "Things" are without properties save "color" which merely stands for "new thing". From this spare beginning, Loreto et al. derive Zipf's Law and Heap's Law (Loreto et al., 2016).

With others we are examining the "Theory of the Adjacent Possible" (TAP) process, described by the following equation (Steel et al., 2020):

$$M_{t+1} = M_t + \sum_{i=1}^{M_t} \alpha^i \binom{M_t}{i} \quad , \quad 0 \leq \alpha \leq 1$$

In this process, there are at any time *t*, Mt "things". At each time step, any subset of the Mt things, 1,2,3,… can be used to create a single new thing. At the next period of time, *t*+1, the system has the initial number of things, Mt, plus the new things created (Eq. 1). More specifically, the probability that any subset can be used to create a new thing decreases monotonically with the size of the subset, 1,2,3,… This is an entirely new equation. Because subsets up to size Mt can be used, as Mt increases this process has remarkable properties: If the process is started with only a few things, the number increases at a glacial pace then explodes suddenly. The continuous process reaches infinity in a finite time, hence has a pole. The discrete process does not reach infinity, but explodes ever more rapidly (Steel et al., 2020).

The TAP process already seems to account for three features of many processes in the universe:

*i.* The increasing number of different molecular species, living species, and technological "tools" in the evolution of the universe, in the evolution of the biosphere, and in human technological evolution in the past 2.6 million years (Steel et al., 2020; Koppl et al., 2018).

*ii.* If we re-interpret Mt as the complexity of the most complex thing produced at time *t*, the same theory seems to account for the gradual then explosive diversification in molecular complexity, species complexity, and tool complexity over time. With respect to tools, 2.6 million years ago we had perhaps 5 to 10 equally simple stone tools. Now we have billions ranging in complexity from needles to the Space Station (Koppl et al., 2018).

*iii.* Every object created in the TAP process has one or more immediate ancestor objects and may have 0, 1, 2 or more immediate "children", then grandchildren and further descendants. TAP predicts a power law descent distribution with a slope of −1.1 to −1.35 depending upon parameters (Steel et al., 2020). The immediate and later descendants of a legal patent can by assessed by the citation of the parental patent. Remarkably, analysis of 3,000,000 patents in the US Patent Office from 1835 to 2010 is a power law slope −1.30 (Koppl et al., 2021). Here the "objects" combined are not material at all, but ideas. Presumably, the descent distribution of actual technologies in the field is also a power law of the same slope.

A single theory, TAP, appears to explain three disparate phenomena (Koppl et al., 2021) suggesting that there may be something fundamentally correct about it.

*iv.* The TAP processes herald a *new "fourth law" of thermodynamics* in the non-ergodic universe (Cortes et al., 2021). If we set $\alpha=1.0$, the TAP process generates all the possibilities, TP. The Total Possible, TP, increases over time. If we set $\alpha<1.0$, the TAP process generates the actualized possibilities, TA. The Actualized Possible, TA, a subset of the total possible TP, increases over time. The ratio of these, R = TP/TA also increases over time. Hence the non-ergodicity of the non-ergodic system, R, increases over time. Thus, also the localization of the system in its non-ergodically expanding phase space, 1/R, becomes greater over time. The lower 1/R, the greater the localization. *Over time, the universe creates an ever-tinier subset of what is now possible at the level of molecules, species and tools. This new law plays a major role in the increasing complexity of the universe* (Kauffman, 2022).

In this new 4th law, the increasing localization of systems is within their ever-expanding phase spaces. The relation of this new 4th law to the famous 2nd law of thermodynamics and statistical mechanics in a prestated and fixed phase space where entropy always tends to increase and localization of the system in its fixed phase space tends to decrease, remains to be clarified (Kauffman, 2022).

It is hopeful that the same abstract theory fits these three distinct phenomena. Both the Loreto-Strogatz Urn Model (Loreto et al., 2016) and the TAP process (Steel et al., 2020) are not ergodic. They are abstract representations of processes that reach into an unprestatable adjacent possible. In these two models "*things transform to things". However, there is no notion of "function".*

## 6 Kantian Wholes provide the missing concept of “function”

Any living cell or organism is a non-equilibrium physical system that is a Kantian Whole that has the property that the Parts exist for and by means of the Whole. This provides a proper concept of “function”.

**1)** Any living cell or multi-celled organism is a Kantian Whole. Humans are Kantian Wholes. We exist as complex things in the non-ergodic universe above the level of 500 Daltons for and by means of our hearts, and loops of Henle. Hearts and loops of Henle exist as complex things in the universe for and by means of us. Thus, we are Kantian Wholes, our Parts do exist for and by means of the us, the Whole.

**2)** The function of a Part is that subset of its causal properties that sustains the Whole. The function of the heart is to pump blood which process sustains the whole organism of which is a member. The existence of Kantian Wholes as complex things in a non-ergodic universe above 500 Daltons affords a clear meaning to “function”. Functions are real in the universe.

**3)** The system achieves Catalytic or Task Closure: All reactions and tasks needing fulfillment are fulfilled by components in the same system.

**4)** The system achieves the newly recognized and powerful property of Constraint Closure (Montévil and Mossio, 2015). Cells literally construct themselves.

## 7 The Evolution of Integrated Functionality: Emergent Creativity

We achieve a new understanding of the almost miraculous emergent self-construction and emergent coherent functional organization of processes in an evolving a biosphere: There is no deductive relation between the different uses of any physical thing, such as a protein in a cell that can evolve to be used to catalyze a reaction, to carry a tension load, or to host a molecular motor on which it walks. Cells physically construct themselves.

Therefore, each molecule and structure in evolving cells and organisms in the biosphere stands *ever-available to be selected, alone or with other things, for indefinite adaptive new uses* such that myriad new adaptations and new physical things such as new proteins arise all the time. *The new uses are not open to deduction from the old uses*.

*Functional integration is always maintained, even as it transforms, because the functional evolution of the Parts must always sustain the functioning Kantian Whole upon which selection acts*. Selection acting upon the Whole determines what “gets to exist” for some time in the non-ergodic biosphere. This is “downward causation”. The explanatory arrows do not point only downward (Weinberg, 1994).

The evolving biosphere really is a propagating adapting construction, not an entailed deduction. This is “sustained functional integrated emergence” in evolving Kantian Wholes. It is the arrival of the fitter.

This is emergence. *Emergence is not engineering*. This radical emergence of a co-evolving biosphere itself emerges only beyond the Newtonian Paradigm. That we are at a third transition in science, beyond Newton’s wonderful paradigm is not a loss, rather it is an invitation to participate in this magical emergence we have not even seen before.

We hardly begin to understand this. An evolving biosphere is a self-constructing, functionally integrated blossoming emergence. This seems also to share a common ground with co-dependent origination (Laumakis, 2008).

An evolving biosphere is a propagating construction, not an entailed deduction.

Hiding behind the equations we write, we cannot see the reality that they hide: The mystery of evolving life. We are of it, not above it.

## 8 Abstract Kantian Wholes as a Formalization of Functionality

An *abstract representation of any Kantian Whole includes both things transforming to things, and things regulating these transformations among things*. This is a form of "*digraph*", see Figure 1. In general, "things" are represented by "*circles*", while transformation among things, "reactions", are represented as "*dots*". Every *circle* "thing" is connected to one or more transformation *dots*. Every transformation *dot* is connected to one or more *circle* "thing". This is a digraph.[2]

In addition, the *"thing regulating a specific transformation among things" is represented by an arrow from the "thing" circle to the transformation dot that the "thing" regulates*. The result is a digraph augmented with 0, 1, or more arrows from each thing to any transformation it regulates.

Figure 1 is a typical example of an abstract Kantian Whole. The Kantian Whole has the property that the last step in the formation of each thing is positively regulated by one or more of the things in the set. In this abstract representation of a Kantian Whole, the Parts really get to exist for and by means of the Whole. In a physical Kantian Whole, the function of a part really is the subset of its indefinitely many causal properties that help sustain the Whole.

Magically, precisely because we have abstracted away any specific properties from a "thing", it can come to be used, hence function, in indefinitely many ways. The "engine block", here abstracted from specific properties, can function as a paper weight and also function to crack open coconuts.

[2] A similar formalism has also been introduced by Robert Rosen Rosen (1972, 1991).

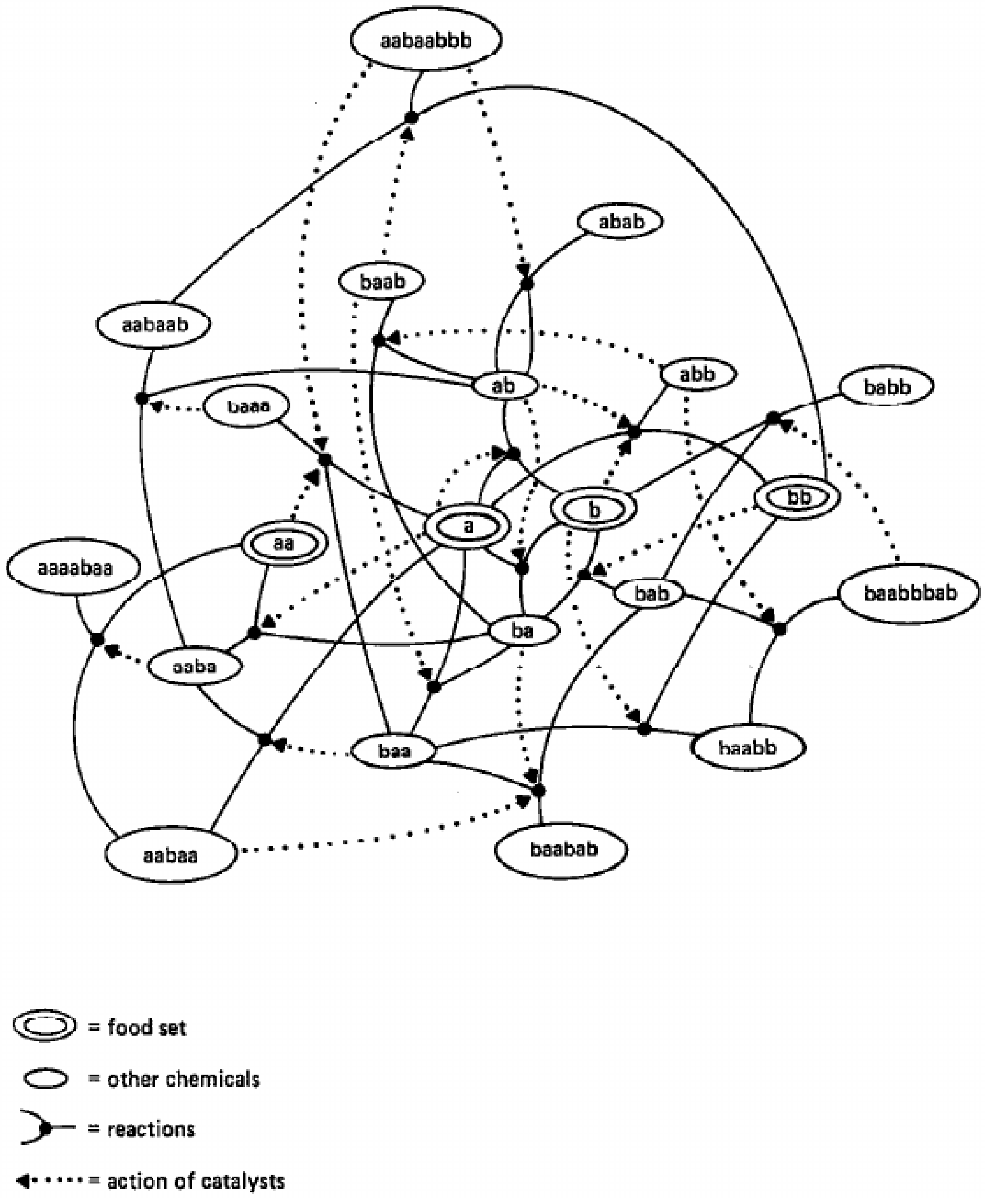

Figure 1: Example of a graph describing an autocatalytic set, taken from (Farmer et al., 1986). Courtesy of the authors.

We have achieved an abstract model of the functional closure of a real physical Kantian Whole, each of whose parts can come to function in ways that cannot be deduced from one another. Again, we achieve this precisely by abstracting away any specific properties of a "thing", the transformations of things to things, and the way things can regulate the transformations. In this abstract representation, a thing can be a molecule, A transforming to B, and regulation can be catalysis of the transformation reaction by C. A thing can be a species in the process of surviving, and the things mediating this process of survival can be the things in the niche of the surviving species. The things can be goods sold in a market, regulated by relevant legal laws. A corporation is a Kantian Whole embedded in the larger Kantian Whole economic and legal world enabled by laws that constrain human activities into the specific human work that sustains the Kantian Whole corporation and its enabling legal laws. By abstracting any properties from a "thing", the indefinite actual uses of any physical thing can be captured. In short, the "physical", "legal", "ideational" character of a "thing" is irrelevant to the abstracted "thing", its transformations and the regulations of transformations among things by things.

If we are precluded from using Set Theory with respect to real things, stones, hammers, and enzymes, we are fully allowed here to use Set Theory with this fully syntactic model of things and their transformations.

# 9 A Statistical Mechanics of Emergence

A Union of the TAP Process and the Evolution of Kantian Wholes: TAP is an abstraction of non-ergodic processes where one or more things can give rise to one new thing. A further step, closer to chemistry, is for one or more things to give rise to one or more things. *To unite the TAP process with the evolution of functionally integrated Kantian Wholes, we have merely to add to TAP that "things" can act on the transformation by which things yield things, to speed or slow the transformation, i.e. catalyze or inhibit the transformation*. Most generally, let K "things" act with some probability, P, on any transformation, X. We can model the effect of the K things on this transformation *by arbitrary Boolean functions on K inputs*. We can explore different rules by which "things" come to act on transformations among things.

*We here achieve for a first time an abstract union of the functionality of Kantian Wholes and the non-ergodic open transformation of things transforming to things by the TAP processes*, or a generalization in which more than one thing can be produced in a transformation. This leads to the formation and evolution of abstract Kantian Wholes with one another within the evolving TAP process as it creates an increasing diversity of "things". The character of the "things" and "transformations" do not matter at all. Again, the things can be molecules reacting and forming a spray of new molecules (Scherer et al., 2017), and perhaps catalyzing or inhibiting those reactions. The things can be goods or services created out of input goods and transformed in factories, and other capital goods, into output goods (Cazzolla Gatti et al., 2020), or legal laws. The transformation can be carried out by human actions legally allowed or forbidden by extant evolving legal laws, as in human action in an economy.

There is a first hint of a *statistical mechanics of emergence in non-ergodic systems*. Whether we consider an evolving chemical system whose molecules transform and regulate the transformations, an ecosystem of species creating and blocking niches for one another, or goods and services in an evolving economy creating and blocking market opportunities for one another, or the evolution of legal systems, we have a new set of conceptual, indeed, mathematical tools.

New questions arise. Over time, how many abstract Kantian Wholes emerge? What are the statistics of their sizes? Do they help or hinder one another? Do they co-evolve? If so, what are the statistical structures of their co-evolving fitness landscapes? Do those landscapes tend to asymptotic forms of criticality (Kauffman, 1995)? Genetic regulatory networks in cells and brains are dynamically critical (Daniels et al., 2018; Villani et al., 2018; Beggs, 2008). Do the Boolean functions in abstract Kantian Wholes evolve to dynamical criticality with small attractors and a maximization of transfer entropy within and between the emergent Kantian Wholes (Li et al., 2019)? Do the systems evolve to tune their own connectivity in some way? Why and how? How many "goods and services" do Kantian Wholes exchange? Does any of this map to molecular and functional trading in microbial communities (Le and Wang, 2020). Does it map small ecosystems? To the entire evolving biosphere? Economy? Legal systems? A bacterium is a Kantian Whole. A eukaryotic cell contains mitochondria which themselves are Kantian Wholes. Thus, a eukaryotic cell is a *second order Kantian Whole enclosing a first order Kantian Whole*. A multi-celled organism is a *third order Kantian Whole* containing second order Kantian Wholes containing first order Kantian Wholes. The ecosystem in our guts and our cells is a *fourth order Kantian Whole* whose parts exist for and by means of the Whole. Probably the entire biosphere is some form of high order nested Kantian Whole. Is an economy a nested set of Kantian Wholes? With what emergent statistical regularities? Might it be possible to study abstract statistical properties of emergence of nested higher order evolving Kantian Wholes in a Statistical Mechanics of Emergence?

## 10 Agency Function Purpose, Teleonomy, and Evolution

The very existence of Kantian Wholes in the non-ergodic universe above the level of atoms allows, as noted, a noncircular definition of "function". The function of a Part is that subset of its causal properties that sustains the Kantian Whole. What of Purpose? As Jacque Monod famously said in "Chance and Necessity" (Monod, 1971), the project of every cell is to create two cells. Living cells are molecular autonomous agents, able to reproduce, perform thermodynamic work cycles, and choose, (Clayton and Kauffman, 2006). To choose, cells must sense the world, evaluate it "good or bad for me", and act (Peil, 2014). This triad is central to the project of every cell and is the root of "value" (Peil, 2014). The same triad is the root of affordances: "goal", "affordance", "means to seize the affordance" (Walsh, 2015). Here is where teleonomy comes into play: organisms act according to their "values", and so their internal goals, and the opportunities they have. The tight, irreducible coupling between goals and affordances emphasizes the prominent role of organism purposiveness in the process of affordance seizing exerted by natural selection, which is central for *Teleonomic Selection* (Corning, 2014). Furthermore, as living organisms are nested Kantian Wholes, teleonomy can be found at every level of organisms and so its impact on evolution.

All this is central to the roles of consciousness is evolution. As we have seen, "seeing an affordance", such as seeing the use of an engine block to crack open coconuts, cannot be deduced. As we argue elsewhere (Kauffman and Roli, 2022) "seeing" complex sequential affordances, as in jury-rigging (Jacob, 1977), cannot be achieved by non-embodied Universal Turing Machine or embodied robots, but can be achieved by us. This suggests that mind is quantum and perhaps that quantum actualization underlies the consciousness that allows us to "see complex affordances". Were this true, it would allow mind to have evolved and played its diverse roles as organisms mutually created more complex worlds with one another over the past 3.7 billion years.

# 11 Conclusions

The 21st Century promises to be the Century of Biology. This embraces of course the explosion of biotechnology, an emergence of 21st Century medicine, and ever deeper analysis of how cells and organisms that now exist "work" as physical systems at molecular, cellular, organism, and ecosystem levels. Here reliance on physics, chemistry, biophysics, biochemistry and molecular biology is essential. The issues are massive in complexity and import. We are in the Era of Systems Biology.

However, we confront the third major transformation in science, following Newton and Quantum Mechanics, the first two transformations. We are forced beyond the wonderful Newtonian Paradigm. There really is no "Final Theory": The diachronic evolution of our or any biosphere is beyond entailing law and beyond any mathematics based on Set Theory.

There may well be biospheres in the universe. Evolving biospheres are immensely creative in ways beyond our knowing or stating. We live forward in face of mystery. This implies that we humans are Of Nature, not Above Nature. Rather than a loss, this is, instead, an enormous invitation. We can try to understand in new ways how our or any biosphere, our global economy, and even our cultures diachronically construct themselves over billions, millions, and hundreds of thousands of years of unprestatable, non-entailed, ever-creative, nonergodic emergence. We are invited to construct a new statistical mechanics of emergence. We come to understand that we really are conscious agents. We are also invited to live responsibly in our shared biosphere.

**Conflicts of interest:**

Right to publish. The authors have no competing obligations.

# References

Arrow, K., & Debreu, G. (1954). Existence of an equilibrium for a competitive economy. *Econometrica: Journal of the Econometric Society, 22(3)*, 265–290.

Ashkenasy, G., Jagasia, R., Yadav, M., & Ghadiri M. (2004). Design of a directed molecular network. *Proceedings of the National Academy of Sciences, 101*(30), 10872–10877.

Aspect, A., Dalibard, J., & Roger, G. (1982). Experimental test of Bell's inequalities using time-varying analyzers. *Physical review letters, 49*(25), 1804–1807.

Atkins, P. (1984). *The second law*. New York, NY: Scientific American Library.

Barve, A., & Wagner, A. (2013). A latent capacity for evolutionary innovation through exaptation in metabolic systems. *Nature 500*(7461), 203–206.

Beggs, J. (2008). The criticality hypothesis: how local cortical networks might optimize information processing. *Philosophical Transactions of the Royal Society A: Mathematical, Physical and Engineering Sciences 366*(1864), 329–343.

Birkhoff, G., & Von Neumann, J. (1936). The logic of quantum mechanics. *Annals of mathematics*, *37(4)*, 823–843.

Cass, D., & Shell, K. (1983). Do sunspots matter? *Journal of political economy 91*(2), 193–227.

Cazzolla Gatti, R., Koppl, R., Fath, B., Kauffman, S., Hordijk, W., & Ulanowicz, R. (2020). On the emergence of ecological and economic niches. *Journal of Bioeconomics 22(2)*, 99–127.

Clayton, P., & Kauffman, S. (2006). On Emergence, Agency, and Organization. *Biology and Philosophy, 21(4)*, 501-521.

Corning, P. A. (2014). Evolution 'on purpose': how behaviour has shaped the evolutionary process. *Biological Journal of the Linnean Society*, *112*(2), 242-260.

Cortês, M., Kauffman, S., Liddle, A. R., & Smolin, L. (2022). Biocosmology: Biology from a cosmological perspective. *arXiv preprint arXiv:2204.09379*.

Daniels, B. C., Kim, H., Moore, D., Zhou, S., Smith, H. B., Karas, B., Kauffman, S., & Walker, S.I. (2018). Criticality distinguishes the ensemble of biological regulatory networks. *Physical review letters*, *121*(13), 138102.
Einstein, A., Podolsky, B., & Rosen, N. (1935). Can quantum-mechanical description of physical reality be considered complete? *Physical review 47*(10), 777.
Farmer, J., Kauffman, S., & Packard, N. (1986). Autocatalytic replication of polymers. *Physica D: Nonlinear Phenomena 22*(1-3), 50–67.
Feynman, R. (1998). *Quantum electrodynamics*, Nashville, TN: Westview Press.
Franzén, T. (2005). *Gödel's theorem: an incomplete guide to its use and abuse*. Boca Raton, FL: CRC Press.
Geanakoplos, J., & Polemarchakis, H. (1986). Existence, regularity and constrained suboptimality of competitive allocations when the asset structure is incomplete. In W. Hell, R. Starr, and D. Starrett (Eds.), *Uncertainty, information and communication: Essays in honor of K.J. Arrow*, Chapter 3, pp. 65–95. Cambridge, UK: Cambridge University Press.
Gibson, J. (1966). *The senses considered as perceptual systems*. Boston, MA: Houghton Mifflin.
Gould, S. J., & Vrba, E.S. (1982). Exaptation—a missing term in the science of form. *Paleobiology*, *8*(1), 4-15.
Grabiner, J. (1983). Who gave you the epsilon? Cauchy and the origins of rigorous calculus. *The American Mathematical Monthly 90*(3), 185–194.
Heisenberg, W. (1958). *Physics and philosophy: The revolution in modern science*. New York, NY: Harper Torchbooks.
Hordijk, W., & Steel, M. (2004). Detecting autocatalytic, self-sustaining sets in chemical reaction systems. *Journal of theoretical biology 227*(4), 451–461.
Jacob, F. (1977). Evolution and Tinkering. *Science New Series 196(4295)*, 1161–1166.
Jech, T. (2006). *Set theory*. New York, NY: Springer.
Joyce, G. (2002). The antiquity of RNA-based evolution. *nature 418*(6894), 214–221.
Kaku, M. (2021). *The God Equation: The Quest for a Theory of Everything*. New York, NY: Doubleday.
Kauffman, S. (1970). Articulation of parts explanation in biology and the rational search for them. In *Topics in the Philosophy of Biology*, 245–263. Springer.
Kauffman, S. (1971). Cellular homeostasis, epigenesis and replication in randomly aggregated macromolecular systems. *Journal of Cybernetics 1*(1), 71–96.
Kauffman, S. (1986). Autocatalytic sets of proteins. *Journal of theoretical biology 119*(1), 1–24.
Kauffman, S. (1995). *At home in the universe: The search for the laws of self-organization and complexity*. Oxford University Press.
Kauffman, S. (2016). *Humanity in a creative universe*. Oxford, UK: Oxford University Press.
Kauffman, S. (2019). *A world beyond physics: the emergence and evolution of life*. Oxford, UK: Oxford University Press.
Kauffman, S. (2022). *arXiv preprint arXiv:2205.09762*.
Kauffman, S. (2020). Answering Schrödinger's "What Is Life? ". *Entropy 22*(8), 815.
Kauffman, S., Jelenfi, D., & Vattay, G. (2020). Theory of chemical evolution of molecule compositions in the universe, in the miller–urey experiment and the mass distribution of interstellar and intergalactic molecules. *Journal of theoretical biology 486*, 110097.
Kauffman, S., & Roli, A. (2021). The world is not a theorem. *Entropy 23*(11), 1467.
Kauffman, S., & Roli, A. (2022). What is consciousness? Artificial intelligence, real intelligence, quantum mind, and qualia. *Submitted. A previous version of the paper is available as arXiv preprint arXiv:2106.15515*.
Koppl, R., Devereaux, A., Herriot, J., & Kauffman, S. (2018). A simple combinatorial model of world economic history. *arXiv preprint arXiv:1811.04502*.
Koppl, R., Devereaux, A., Valverde, S., Solé, R., Kauffman, S., & Herriot, J. (2021, May 30). Explaining technology. *SSRN papers, https://papers.ssrn.com/sol3/papers.cfm? abstract_id=3856338*.
Kvenvolden, K., Lawless, J., Pering, K., Peterson, E., Flores, J., Ponnamperuma, C., Kaplan, I., & Moore, C. (1970). Evidence for extraterrestrial amino-acids and hydrocarbons in the murchison meteorite. *Nature 228*(5275), 923–926.
Lancet, D., Zidovetzki, R., & Markovitch, O. (2018). Systems protobiology: origin of life in lipid catalytic networks. *Journal of The Royal Society Interface 15*(144), 20180159.
Laumakis, S. (2008). Interdependent arising. *Cambridge Introductions to Philosophy*, pp. 105–124. Cambridge University Press.

Le, M., & Wang, D. (2020). Structure and membership of gut microbial communities in multiple fish cryptic species under potential migratory effects. *Scientific reports 10*(1), 1–12.
Lehman, N., & Kauffman, S. (2021). Constraint closure drove major transitions in the origins of life. *Entropy 23*(1), 105.
Li, M., Han, Y., Aburn, M., Breakspear, M., Poldrack, R., Shine, J., & Lizier, J. (2019). Transitions in information processing dynamics at the whole-brain network level are driven by alterations in neural gain. *PLoS computational biology 15*(10), e1006957.
Lincoln, T., & Joyce, G. (2009). Self-sustained replication of an RNA enzyme. *Science 323*(5918), 1229–1232.
Longo, G. (2019). Interfaces of incompleteness. In G. Minati, M. Abram, and E. Pessa (Eds.), *Systemics of Incompleteness and Quasi-systems*, pp. 3–55. Springer.
Longo, G., Montévil, M., & Kauffman, S. (2012). No entailing laws, but enablement in the evolution of the biosphere. In *Proceedings of GECCO 2012 – The 14th Genetic and Evolutionary Computation Conference*, pp. 1379–1392.
Loreto, V., Servedio, V., Strogatz, S., & Tria, F. (2016). Dynamics on expanding spaces: modeling the emergence of novelties. In *Creativity and universality in language*, pp. 59–83. Springer.
Monod, J. (1971) *Chance and Necessity: An Essay on the Natural Philosophy of Modern Biology*. New York, NY:Alfred A. Knopf.
Montévil, M., & Mossio, M. (2015). Biological organisation as closure of constraints. *Journal of theoretical biology 372*, 179–191.
Moore, G. (2012). *Zermelo's axiom of choice: Its origins, development, and influence*. North Chelmsford, MA: Courier Corporation.
Peano, G. (1889). *Arithmetices principia: Nova methodo exposita*. Turin, Italy: Fratres Bocca.
Peil, K.T. (2014) Emotion: the self-regulatory sense. *Global advances in health and medicine, 3(2)*, 80–108.
Persons IV, W., & Currie, P. (2015). Bristles before down: a new perspective on the functional origin of feathers. *Evolution 69*(4), 857–862.
Planck, M. (1901). On the law of distribution of energy in the normal spectrum. *Annalen der physik 4,* 553.
Potter, M. (2004). *Set theory and its philosophy: A critical introduction*. Oxford, UK: Clarendon Press.
Prum, R., & Brush, A. (2002). The evolutionary origin and diversification of feathers. *The Quarterly review of biology 77*(3), 261–295.
Robinson, A. (2016). *Non-standard analysis*. Princeton, NJ: Princeton University Press.
Rosen, R. (1972). Some relational cell models: the metabolism-repair systems. In *Foundations of mathematical biology*, Chapter 4, pp. 217–253. Elsevier.
Rosen, R. (1991). *Life itself: a comprehensive inquiry into the nature, origin, and fabrication of life*. New York, NY: Columbia University Press.
Russell, B., & Whitehead, A. (1973). *Principia mathematica*. Cambridge, UK: Cambridge University Press.
Scherer, S., Wollrab, E., Codutti, L., Carlomagno, T., da Costa, S., Volkmer, A., Bronja, A., Schmitz, O., & Ott, A. (2017). Chemical analysis of a "Miller-Type" complex prebiotic broth - part II: Gas, oil, water and the oil/water-interface. *Origins of life and evolution of the biosphere: the journal of the International Society for the Study of the Origin of Life 47*(4), 381–403.
Smolin, L. (2013). *Time reborn: From the crisis in physics to the future of the universe*. London, UK: Penguin Books.
Steel, M., Hordijk, W., & Kauffman, S. (2020). Dynamics of a birth–death process based on combinatorial innovation. *Journal of theoretical biology 491*, 110187.
Svirezhev, Y. (2008). Nonlinearities in mathematical ecology: Phenomena and models: Would we live in volterra's world? *Ecological Modelling 216*(2), 89–101.
Vaidya, N., Manapat, M., Chen, I., Xulvi-Brunet, R., Hayden, E., & Lehman, N. (2012). Spontaneous network formation among cooperative RNA replicators. *Nature 491*(7422), 72–77.
Vasas, V., Fernando, C., Santos, M., Kauffman, S., & Szathmáry, E. (2012). Evolution before genes. *Biology direct 7*(1), 1–14.
Villani, M., La Rocca, L., Kauffman, S., & Serra, R. (2018). Dynamical criticality in gene regulatory networks. *Complexity, 980636.*
von Kiedrowski, G. (1986). A self-replicating hexadeoxynucleotide. *Angewandte Chemie International Edition in English 25(10)*, 932–935.
Walsh, D. (2015). *Organisms, agency, and evolution*. Cambridge, UK: Cambridge University Press.
Weinberg, S. (1994). *Dreams of a final theory*. New York, NY: Vintage.

Xavier, J., Hordijk, W., Kauffman, S., Steel, M., & Martin, W. (2020). Autocatalytic chemical networks at the origin of metabolism. *Proceedings of the Royal Society B 287*(1922), 20192377.